\documentclass[letterpaper,preprint,prl,aps,superscriptaddress,floatfix]{revtex4}
\usepackage[latin1]{inputenc}
\usepackage{bm}
\usepackage{multirow,amssymb,amsbsy,amsmath}
\usepackage{graphicx}
\usepackage{verbatim}
\usepackage{color}

\makeatletter
\usepackage{pifont}
\makeatother

\usepackage{graphicx}
\usepackage{dcolumn}
\usepackage{bm}
\usepackage{ifthen}
\usepackage{booktabs}
\usepackage{nicefrac}
\usepackage{float}
\usepackage{bbm}
\usepackage{amsmath}

\begin{document}

\title{Experimental Optimal Verification of Entangled States using Local Measurements}
\author{Wen-Hao Zhang}
\author{Zhe Chen}
\author{Xing-Xiang Peng}
\author{Xiao-Ye Xu}
\email{xuxiaoye@ustc.edu.cn}
\author{Peng Yin}
\author{Shang Yu}
\author{Xiang-Jun Ye}
\author{Yong-Jian Han}
\author{Jin-Shi Xu}
\author{Geng Chen}
\email{chengeng@ustc.edu.cn}
\author{Chuan-Feng Li}
\email{cfli@ustc.edu.cn}
\author{Guang-Can Guo}
\affiliation{CAS Key Laboratory of Quantum Information, University of Science and Technology of China, Hefei, Anhui 230026, China.}
\affiliation{CAS Center For Excellence in Quantum Information and Quantum Physics, University of Science and Technology of China, Hefei, Anhui 230026, China.}
\date{\today}

\begin{abstract}
The initialization of a quantum system into a certain state is a crucial aspect of quantum information science.
While a variety of measurement strategies have been developed to characterize how well the system is initialized, for a given one, there is in general a trade-off between its efficiency and the accessible information of the quantum state.
Conventional quantum state tomography can characterize unknown states by reconstructing the density matrix; however, its exponentially expensive postprocessing is likely to produce a deviate result. Alternatively, quantum state verification provides a technique to quantify the prepared state with significantly fewer measurements, especially for quantum entangled states. Here, we experimentally implement an optimal verification of entangled states with local measurements, where the estimated infidelity is inversely proportional to the number of measurements. The utilized strategy is tolerant of the impurity of realistic states, hence being highly robust in a practical sense. Even more valuable, our method only requires local measurements, which incurs only a small constant-factor ($<2.5$) penalty compared to the globally optimal strategy requiring nonlocal measurements.

\end{abstract}

\maketitle

\section{Introduction}
Quantum information is a field where the information is encoded into quantum states, and taking advantage of the ``quantumness" of these states, we can perform more efficient computations \cite{Shor} and more secure cryptography \cite{Bennett} compared to their classical counterparts. Basically, there are two preconditions for the achievement of these breakthroughs. On one hand, we need to initialize quantum systems into the target quantum states reliably. On the other hand, we have to develop precise and efficient techniques to characterise the prepared states. A variety of techniques have been developed for quantum state characterization, each of which is designed and optimized along the specific scenario. Quantum state tomography (QST) \cite{James} is designed for characterizing a completely unknown state, and the reconstructed density matrix can provide the full information of the quantum state. However, the conventional tomographic reconstruction of a state is an exponentially time-consuming and computationally difficult process. Adaptive QST relies on solving an optimization problem using previous results, hence to be computationally expensive as standard QST \cite{Mahler,Qi,Chapman,Yang}. With the prior knowledge that the measured states are within some special categories of states, compressed sensing \cite{Flammia1,Gross} and matrix product state tomography \cite{Cramer} can achieve a significantly higher efficiency compared to QST. Entanglement witnesses extract only partial information about the state with far fewer measurements \cite{Toth1,Toth2}. In recent years, a device-independent method, namely, self-testing, has provided a guarantee of the lowest possible fidelity as the figure of merit of the tested devices \cite{Zhang1,Zhang2,Zhang3,Jed,Coladangelo}.

From a practical point of view, in many important scenarios, we merely need to know how well the system is prepared into a target state. In other words, we need to certify that the on-hand quantum device, designed to produce a particular quantum state, does indeed produce states that are sufficiently close to the target states.
Apparently, QST can be applied in these scenarios and provide complete information about the actual states; however, its exponential complexity prevents its application to large-scale systems.
Alternatively, quantum state verification (QSV) represents a technique to overcome this shortage at substantially lower complexity. Concretely, QSV is implemented by performing a sufficient amount of measurements on the output states, and eventually the verifier reaches a conclusion from the statistical analysis, such as: ``the device outputs copies of a state that has at least $1-\epsilon$ fidelity with the target, with $1-\delta$ confidence". This conclusion can be read into two stages: first, the verifier certifies that the produced states are within a ball of $\epsilon$ radius of the correct state; second, this conclusion might be incorrect with a probability of $\delta$.

Intuitively, optimizing the performance of a specific verification strategy can be realized by minimizing $\epsilon$ or $\delta$ with a given number of measurements $n$. Rigorously, this performance is quantified as the scaling of $\epsilon$ or $\delta$ with $n$. In the realm of physical parameter estimation, the precision is normally standard-quantum-limited, which is proportional to $1/\sqrt{n}$. Heisenberg-scaling metrology can achieve $1/n$ scaling; however, its realization generally requires precious quantum resources, e.g., quantum entanglement \cite{Chen1,Chen2}. Quantum state measurement, also faces a similar situation; it is difficult to attain the $1/n$ scaling with normal methods \cite{Sugiyama,Flammia2}.
For the QSV of multi-partite quantum system, $1/n$ scaling can be attained via entangled measurement \cite{Pallister}, which is also a rare quantum resource and difficult to implement in experiment \cite{Gisin}. Recently, a theoretical breakthrough suggests that it is also possible to realize $1/n$ scaling in QSV with local measurements \cite{Pallister}. However, when applying their strategy to a real QSV experiment, impurity of realistic states is likely to produce a rejection outcome within a very limited number of measurements, which prevents the QSV from a valid conclusion. In this work, we experimentally investigate this strategy on a series of two-qubit entangled states with impurities as low as 0.005, and the statistical analysis is modified to be tolerant of single rejections. These features make our scheme robust to state imperfections, and eventually achieve a remarkable improvement in efficiency over previous works. Our results clearly indicate that an optimal verification of entangled states has been realized, with only nonadaptive and noncollective local measurements.

\section{Results}
\begin{figure}[htbp]
	\centering
	\includegraphics[width=6in]{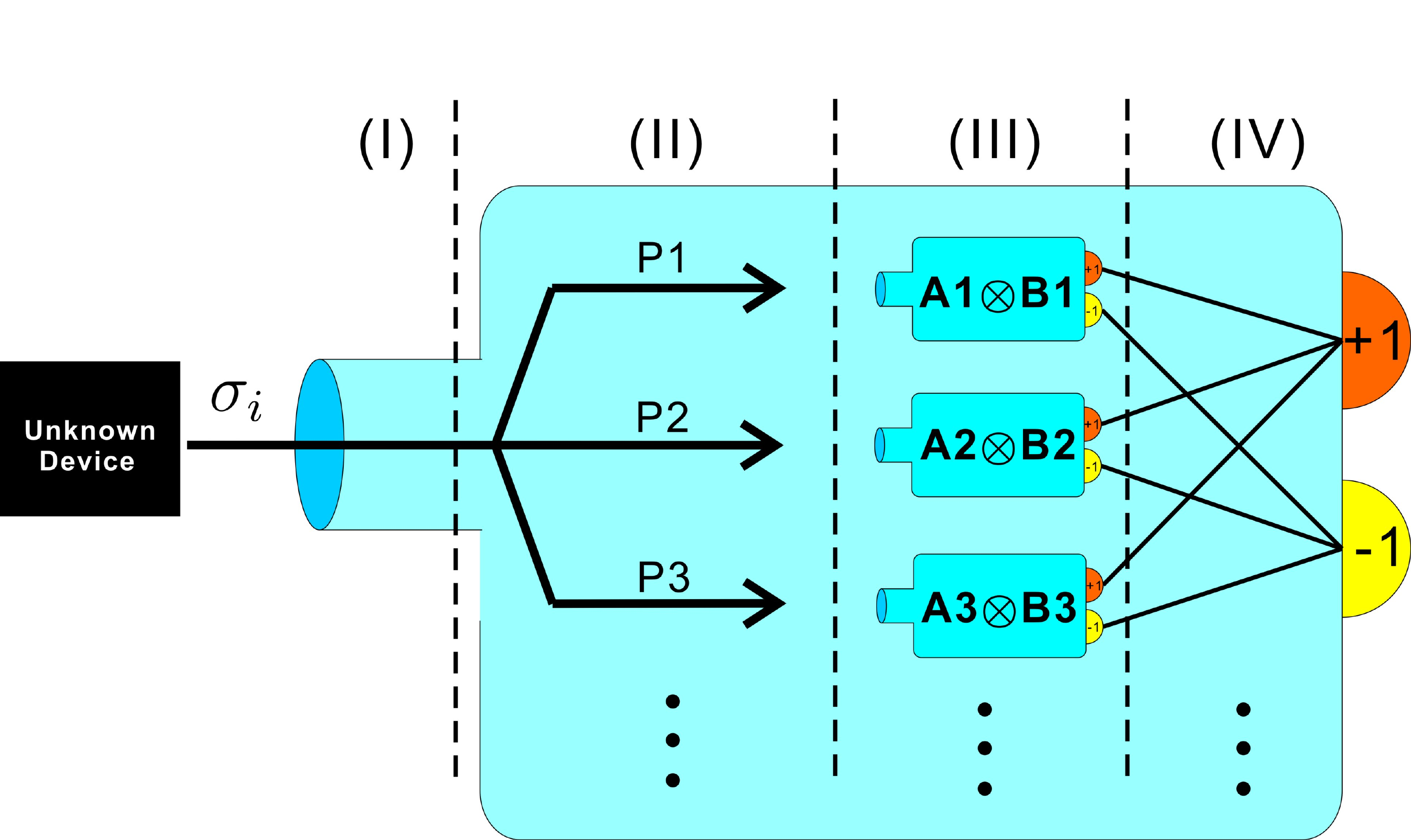}
	\caption{\textbf{Diagram of the QSV of two-qubit entangled states with local measurements.} The figure represents the general procedure to verify whether an unknown source generates a pure target state. The generated states $\sigma_1, \sigma_2, \sigma_3 ,...,\sigma_n$, are projected with randomly selected measurement settings $P_{i}$, which constitute the locally optimal strategy. The binary outcomes are recorded for each measurement, and after completing $n$ measurements, the outcomes are coarse grained as accepted (``-1'') or rejected (``+1'') events. From these analyses, the verifier attempts to ascertain the largest possible distance between the actual and target states, which can be characterized by the infidelity $\epsilon$, with a confidence level of $1-\delta$.}
	\label{strategy}
\end{figure}

\textbf{Robust and optimal QSV proposal with local measurements.} We consider QSV in a real experimental situation. The quantum device produces a series of quantum states  $\sigma_1, \sigma_2, \sigma_3 ,...,\sigma_n$, which are all supposed to be $|\Psi\rangle$ under the assumption of independent and identical distribution. In practice, these states may deviate from $|\Psi\rangle$ or even be different from each other. However, it is still reasonable to expect that all these states are close to $|\Psi\rangle$ with the following fidelity:
\begin{equation}
\label{prior information}
\langle\Psi| \sigma_i |\Psi\rangle \ge1-\epsilon \ (i=1,2,...,n).
\end{equation}
The task for the verifier is to certify that this is the true case and minimize the uncertainty $\epsilon$ by performing a finite number of measurements. Especially, these measurements are expected to be local, nonadaptive, and noncollective. Obviously, for a pure target state, the globally optimal strategy is to project $\sigma_i$ to the target state $|\Psi\rangle$ and its orthogonal state $|\Psi^\perp\rangle$. Because a verification procedure consists of a finite number of measurements and the outcome of each measurement is inevitably probabilistic, the states dissatisfying Eq.\,\ref{prior information} still have a certain probability to pass the verification. The confidence level of giving the conclusion is defined to be $1-\delta$ with $\delta$ to be a small value standing for the probability giving wrong conclusion. With a globally optimal strategy, the minimum number of measurements $n$ required to achieve a certain value of $\epsilon$ and $\delta$ is then given by
\begin{equation}
\label{global optimal}
n^{\text{opt}}_{\text{global}} =-\frac{\ln(\delta^{-1})}{\ln(1-\epsilon)} \overset{\epsilon\rightarrow 0}{\approx}\frac{\ln(\delta^{-1})}{\epsilon}.
\end{equation}
It can be seen that a $1/n$ scaling can be realized for the infidelity $\epsilon$ with a given confidence level. The coefficient is $\ln(\delta^{-1})$, which remains a small constant for a high confidence level. Globally optimal scaling allows for a high verification efficiency; however, entangled measurements are intricate to implement and are regarded as precious resources in quantum information processing \cite{lutk,Vaidman,Calsamiglia,Ewert}. In contrast, local measurements are more feasible in a practical sense. A recent theoretical breakthrough showed that it is possible to achieve a $1/n$ scaling in the verification of two-qubit pure entangled states $|\varphi(\theta)\rangle=\cos\theta|01\rangle-\sin\theta|10\rangle$, while simply with a few local measurements settings \cite{Pallister}. Without loss of generality, considering the task to certify the singlet state $\frac{1}{\sqrt{2}}(|01\rangle-|10\rangle )$, one locally optimal strategy utilizing three measurement settings is written as:
\begin{equation}
\label{singlet1}
\begin{split}
\Omega^{\text{opt}}=\frac{1}{3}(P_{XX}^-+P_{YY}^-+P_{ZZ}^-), \\
\end{split}
\end{equation}
where $P_{XX}^-$ is the projector onto the negative eigen-subspace of the tensor product of Pauli matrices XX (and likewise for $P_{YY}^-$ and $P_{ZZ}^-$).

For each state from the device, a measurement strategy $\Omega$ is applied and the verifier accepts the outcome ``-1" or rejects the outcome ``+1". In an ideal case, where these states are perfectly pure and all his received outcomes are ``-1", the relation between the number of measurements $n$, the fidelity $1-\epsilon$ and the confidence level $1-\delta$ can be written as:
\begin{equation}
\label{local optimal}
n^{\text{opt}}_{\text{local}}{=}  \tfrac{\ln\delta^{-1}}{\ln[1-f(\theta)*\epsilon]^{-1}}\overset{\epsilon\rightarrow 0}{\approx} \tfrac{1}{f(\theta)*\epsilon} \ln\delta^{-1}.
\end{equation}
Evidently, a $1/n$ scaling of $\epsilon$ with $n$ is predicted by this equation, which approaches the globally optimal strategy and slower
by a factor of $f(\theta)$. For the singlet state $f(\theta)$ is calculated to be 2/3, while it depends on the specific strategy for other $|\varphi(\theta)\rangle$. This equation can be read in different ways according to the given task. If the verifier wants to certify the state up to certain fidelity $1-\epsilon$ and confidence level $1-\delta$, Eq.\,\ref{local optimal} gives the lowest number of measurements that must be performed. From another point of view, if the verifier has performed $n$ measurement with all the outcomes being ``-1", the achieved fidelity $1-\epsilon$ and confidence level $1-\delta$ can also be identified from Eq.\,\ref{local optimal}.

As all these profits are built on the premise that all the outcomes should be ``-1", a single appearance of ``+1" will cease the verification without any valid statement. In practice, the imperfections in the experiment will unavoidably yield a certain probability to observe ``+1" for each measurement. Although single rejection probability is small, the probability that all the outcomes are accepted decreases exponentially with the number of measurements; therefore, the current QSV strategy is severely limited for experimental implementation.

A modified strategy is thus developed here by considering the proportion of accepted outcomes, which is tolerant of a certain degree of state imperfections. Quantitatively, we have the corollary that if $\langle\Psi| \sigma_i |\Psi\rangle \leq 1-\epsilon$ for all the measured states, the probability for each outcome to be accepted is smaller than $1-f(\theta)*\epsilon$. As a result, in the case that the verifier observes an accepted probability $ p\ge 1-f(\theta)*\epsilon$, it should be concluded that the actual state satisfies Eq.\,\ref{prior information} with confidence level to be $1-\delta$, where $\epsilon$ and $\delta$ are calculated from the equation \cite{Dimic}
\begin{equation}
\label{tensorproduct}
\delta \leq e^{-D(\frac{m}{n}||1-f(\theta)*\epsilon) n},
\end{equation}
with
\begin{equation}
\label{div}
\begin{split}
&D(x||y)=x \log \frac{x}{y}+(1-x) \log \frac{1-x}{1-y},      \\
\end{split}
\end{equation}
and $m$ is the number of accepted outcomes among $n$ measurements. Benefiting from this modification, single rejections only lead to a decrease in the expected fidelity or confidence. If only the final accepted probability $ p\ge 1-f(\theta)*\epsilon$, the verifier can still give a valid statement about the distance between the actual and target states.

\begin{figure}[htbp]
\centering
\includegraphics[width=6in]{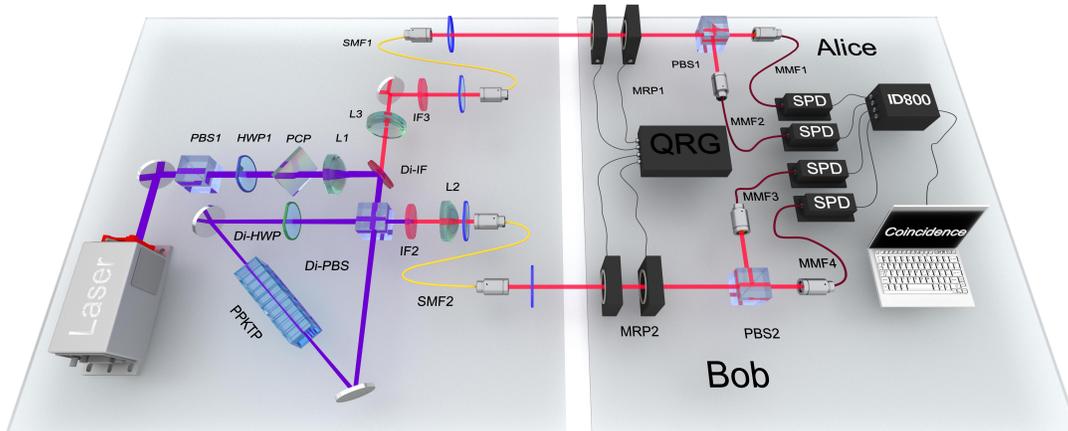}
\caption{\textbf{Experimental setup.} The setup includes an entanglement source and a measurement apparatus for QSV. Entangled photon pairs are generated by pumping a periodically poled KTP (PPKTP) crystal in a polarization Sagnac interferometer (SI). A QRG is connected to two sets of motorized rotated plates (MRP), each of which includes a half-wave plate (HWP), a quarter-wave plate (QWP) and a polarization beam splitter (PBS). The output signal from QRG determines which setting is selected in each measurement. Each pair of entangled states will be detected with two single photon detectors (SPDs). The coincidence is recorded and analyzed by an ID800 (ID Quantique). Di - dichroic, MMF - multi-mode fiber, IF - interference filter, SMF - single mode fiber, PCP - phase compensation plate, L - lens. }
\label{setup}
\end{figure}

\textbf{Experimental Implementation of QSV.} The proposed QSV is performed with the setup shown in Fig.2, which mainly consists of an entangled photon-pair source and a random sampling measurement apparatus. In the first part, tunable two-qubit entangled states are prepared by pumping a nonlinear crystal placed into a phase-stable Sagnac interferometer (SI) (see Method Section for details). Polarization-entangled photon pairs are generated in the state $\cos\theta|HV\rangle - \sin\theta|VH\rangle$ ($H$ and $V$ denote the horizontally and vertically polarized components, respectively) and $\theta$ is controlled by the pumping polarization. A maximally entangled singlet state, partially entangled state and product state can be generated when $\theta=45^\circ, 30^\circ $ and $0^\circ$, respectively. The projective measurements on both sides, which are jointly decided by the same quantum random generator (QRG), are randomly performed. The randomness here means the measurements are unknown to the entangled photon pairs until they are measured.

\begin{figure}[htbp]
\centering
\includegraphics[width=6.5in]{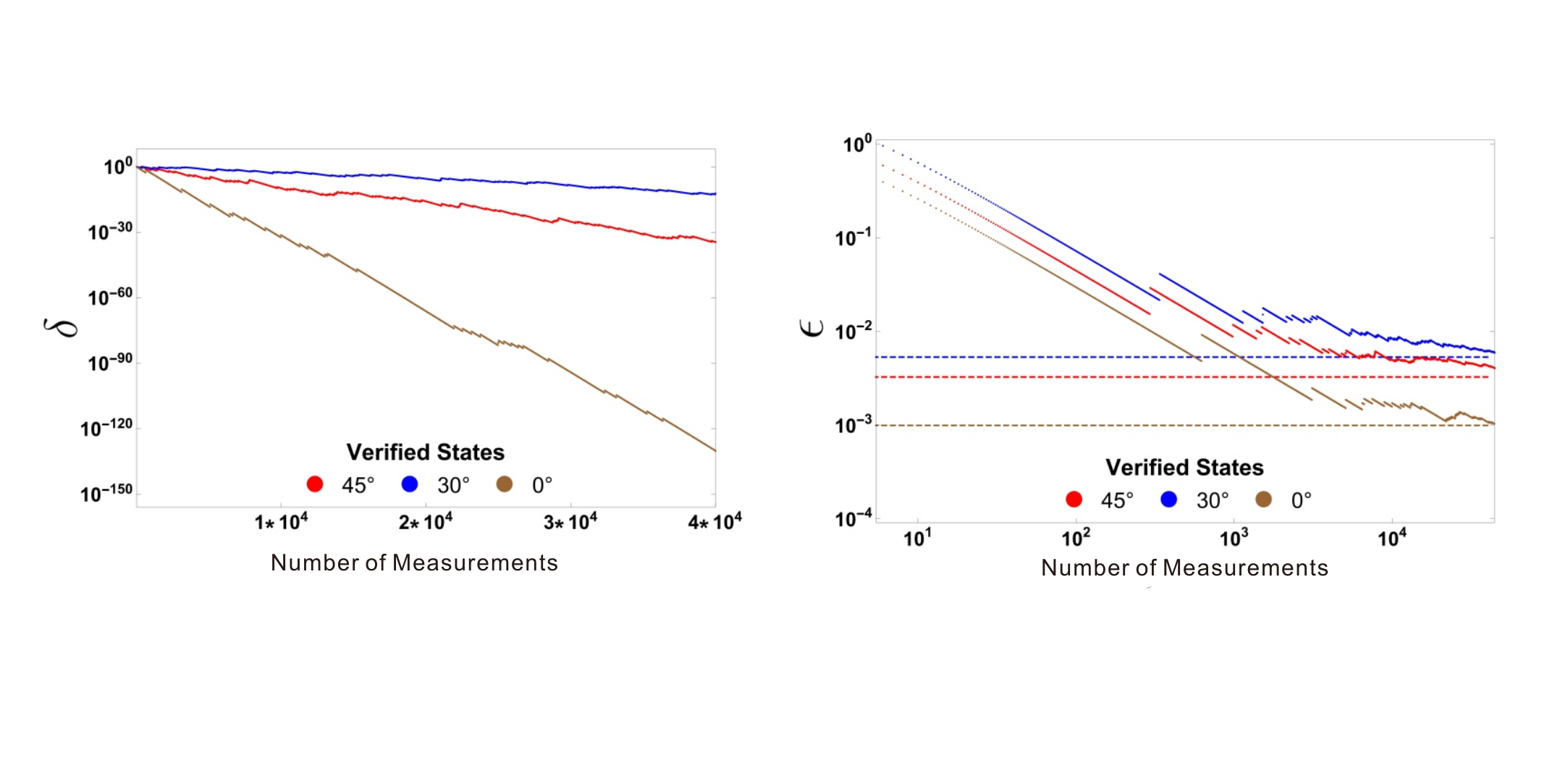}
\caption{\textbf{Experiment results of a single trial of QSV with 40000 measurements.} We choose three representative pure two-qubit entangled states, namely, singlet states, two-qubit partially entangled states and product states with $\theta=45^\circ, 30^\circ$ and $0^\circ$. \textbf{(a)} When $\epsilon$ is set to 0.01, $\delta$ is log plotted versus the number of measurements.  \textbf{(b)} When $\delta$ is set to 0.05, $\epsilon$ is log-log plotted versus the number of measurements. In the log-log plot, the predicted linearity is broken up into discrete short lines due to the slight deviation of the actual states from ideal target states. The dashed horizontal lines represent the tomography results obtained from approximately $10^6$ samples.}
\label{2qubit}
\end{figure}

In the experiment, we perform random projective measurements for quantum states with $\theta=45^\circ, 30^\circ $ and $0^\circ$. For the maximally entangled singlet state with $\theta=45^\circ$, the three measurement settings are selected as those in Eq. (\ref{singlet1}). For a partially entangled state with $\theta=30^\circ$, the proposed optimal strategy can be found in the Method Section. The situation changes for the product state with $\theta=0^\circ$, for which the locally optimal strategy is consistent with the globally optimal strategy, and thus, we only need to project the product state onto itself and its orthogonal complementary space. For all these three states, we perform 40000 measurements in a single trial, and the values of $\delta$ and $\epsilon$ for each measurement are calculated from Eq. (\ref{tensorproduct}), as shown in Fig. 3. In Fig. 3 (a), $\delta$ is calculated with the infidelity set as $\epsilon=0.01$ for all the three tested states. Within 1000 measurements, $\delta$ rapidly approaches 0, meaning that a near-unity degree of confidence level can be achieved with a pretty high efficiency. Alternatively, we can set the confidence level $1-\delta$ to be 0.95 and calculate $\epsilon$. Fig. 3(b) shows that for all the three states the estimated $\epsilon$ descends below 0.01 after 40000 measurements. For both analysis methods, random rejections occur and cause abrupt increases in the estimated values of $\delta$ and $\epsilon$. As a result, neither $\delta$ nor $\epsilon$ can persistently descend with increasing measurements as predicted by Eq. (\ref{local optimal}). One example is the verified infidelity for the singlet state, which approximates the infidelity value from the tomography study after 40000 measurements.

In the study of the scaling of the infidelity with the increasing number of measurements, the results of a single trial of QSV are not convincing because it is too probabilistic, considering the random rejections caused by imperfect realistic states. In order to study the true scaling behavior, we perform 50 trials of verification and average the results ranging from 20 to 80 measurements, as shown in Fig. 4. Due to the high quality of the prepared states, the rejection event in this range is rare, and thus, the exhibited scaling can be regarded as the true performance of the verification proposal. The line for the product state with $\theta=0^\circ$ almost overlaps with that of the globally optimal strategy, which is predicted by Eq. (\ref{global optimal}). For the two entangled states with $\theta=30^\circ$ and $45^\circ$, the observed scaling is also approximately parallel to the globally optimal line, indicating that a $1/n$ scaling verification is achieved in our experiment. The use of local measurement settings only incurs a constant-factor penalty over the globally optimal strategy, which are $2+\frac{\sqrt{3}}{4}$ and 1.5 for $\theta=30^\circ$ and $45^\circ$, respectively.

\begin{figure}[htbp]
\centering
\includegraphics[width=5.5in]{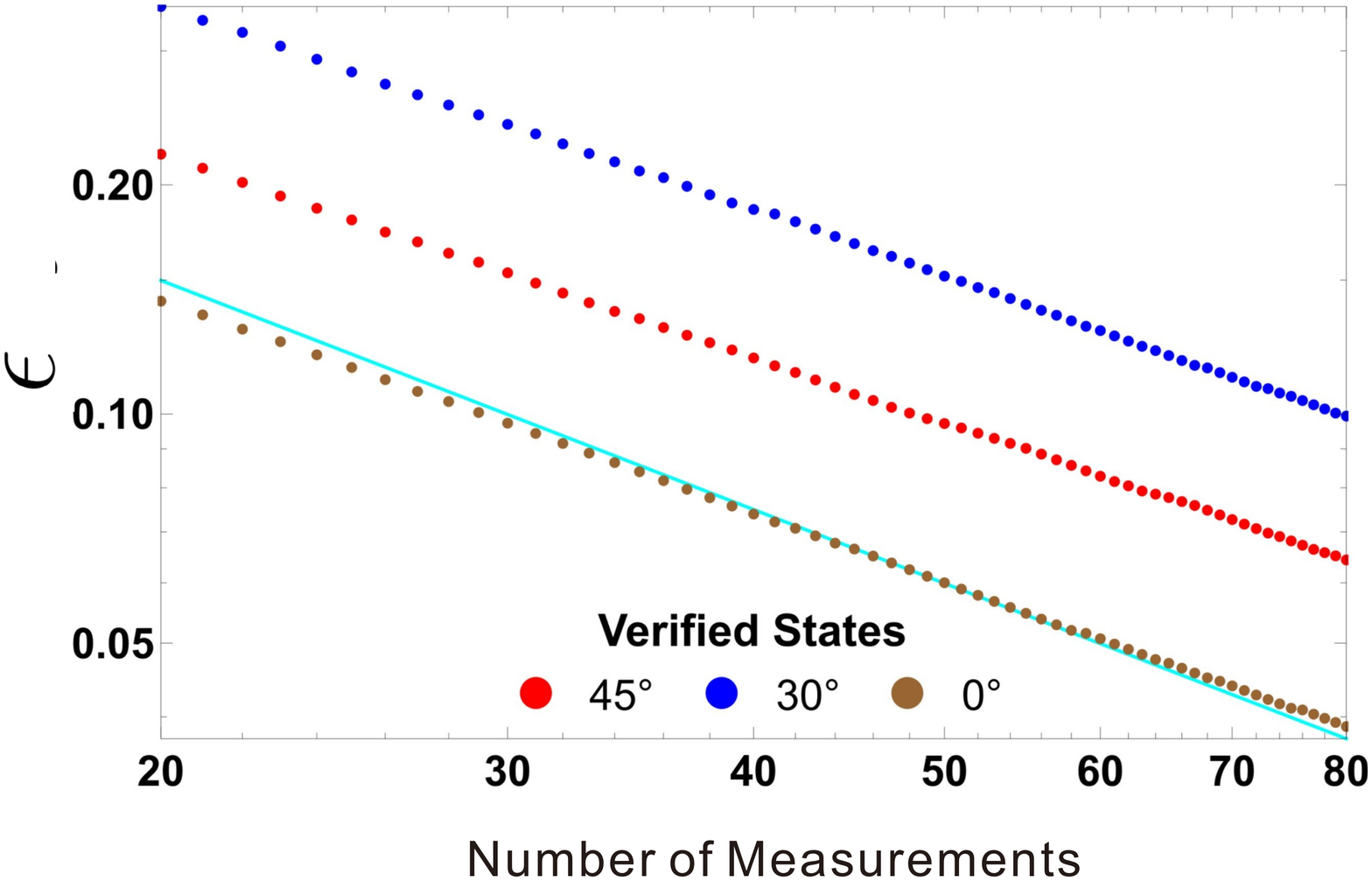}
\caption{\textbf{Average results of $\epsilon$ from 50 trials of QSV.}  We repeat the verification for 50 trials and average the results to study the true scaling of the utilized strategy. Three states with $\theta=45^\circ, 30^\circ $ and $0^\circ$ are verified. The solid cyan  line stands for the globally optimal scaling, which is proportional to $1/n$. The figure is plotted in log-log coordinates, where parallelism represents the same scaling.  }
\label{2qudits tomography}
\end{figure}

\section{Discussion}
Given a confidence level, the QSV strategy utilized here directly accesses the fidelity of the actual state to the target state. Although its description cannot be as comprehensive as that of the point estimation such as state tomography, it achieves superior performance in in three aspects: first, QSV enables much less complex postprocessing and higher efficiency than state tomography if the verifier is only concerned about the fidelity; second, for standard quantum state tomography, maximum likelihood is used to obtain physical estimates for quantum states, which is strongly biased given a small quantity of samples\cite{Schwemmer}; third, the point estimation results may not accurately describe the actually prepared states, because the tomographic reconstruction is given probabilistically and it is difficult to estimate the distance between the target and actual states. Previous works show that adaptive measurements or collective measurements help in achieving better results in tomography. The adaptive measurements require that the experimental setup be able to perform feedback control and varying measurements, which are not easily calibrated in practice \cite{Okamoto}. The apparatus to implement collective measurements \cite{Hou}, such as quantum walk circuits, is also intricate because multi-path interference is difficult to generate and maintain in practical experiment.

To summarize, we experimentally realize an optimal QSV, which is easy to implement and robust to realistic imperfections. The exhibited 1/n scaling results from the strategy itself, without entangled or adaptive measurements. Our results have clear implications for many quantum measurement tasks, and may be used as a firm basis for subsequent work on more complex quantum systems.

\section{Materials and Methods}
\textbf{Generation of entangled photon pairs.}
Concretely, a 405.4 nm single mode laser is used to pump a 5mm long bulk type-II nonlinear periodically poled potassium titanyl phosphate (PPKTP) nonlinear crystal placed into a phase-stable SI to produce polarization entangled photon pairs at 810.8 nm. A PBS followed by a HWP and a PCP are used to control the polarization mode of the pump beam. The lens before and after the SI are used to focus the pump light and collimate the entangled photons, respectively. The interferometer is composed of two high reflective and polarization-maintained mirrors, a Di-HWP and a Di-PBS. ``Di" here means it works for both 405.4 nm and 810.8 nm. The Di-HWP flips the polarization of passing photons, such that the type-II PPKTP can be pumped by the same horizontal light from both clockwise and counterclockwise directions. Di-IF and LPF (Long pass filter) are used to remove the pump beam light. BPF (band pass filter) and SMF are used for spectral and spatial filtering, which can significantly increase the fidelity of entangled states.

The whole setup, in particular the PPKTP, is sensitive to temperature fluctuations. Placing the PPKTP on a temperature controller ($\pm0.002^\circ $C stability) and sealing the SI with an acrylic box would be helpful for improving temperature stability.

Tomography is performed on all the three states with a large amount of samples ($>10^6$) to yield a reliable reference for verification results.

\textbf{Implementation of QSV.}
 The power of the laser is attenuated to decrease the generation rate of entangled photon pairs. As a result, for each randomly selected measurement, with a near-unity probability, there is at most only one coincidence to be recorded. The verifier accepts when Alice's outcome is opposite to that of Bob's; otherwise it rejects.

For partially pure entangled states $|\Psi\rangle=\cos\theta|01\rangle-\sin\theta|10\rangle$ ($\theta\in(0^\circ,90^\circ)$ and $\theta\ne45^\circ$),  the locally optimal strategy involves four measurement settings:
\begin{equation}
\label{partially entangled state1}
\begin{split}
\Omega^{opt}=\frac{2-sin\ 2\theta}{4+sin\ 2\theta}P_{ZZ}^-+\frac{2(1+sin\ \theta)}{3(4+sin\ \theta)}\sum_{k=1}^3 (I-|\phi_k\rangle\langle\phi_k|). \\
\end{split}
\end{equation}
where
\begin{equation}
\label{partially entangled state2}
\begin{split}
&P_{ZZ}^-=|01\rangle\langle01|+ |10\rangle\langle10|,                  \\
&|\phi_1\rangle=\left(\frac{1}{\sqrt{1+tan\ \theta}}|0\rangle+\frac{e^{\frac{2\pi i}{3}}}{\sqrt{1+cot\ \theta}}|1\rangle\right) \otimes \left(\frac{1}{\sqrt{1+cot\ \theta}}|0\rangle+\frac{e^{\frac{2\pi i}{3}}}{\sqrt{1+tan\ \theta}}|1\rangle\right) ,                  \\
&|\phi_2\rangle=\left(\frac{1}{\sqrt{1+tan\ \theta}}|0\rangle+\frac{e^{\frac{4\pi i}{3}}}{\sqrt{1+cot\ \theta}}|1\rangle\right) \otimes \left(\frac{1}{\sqrt{1+cot\ \theta}}|0\rangle+\frac{e^{\frac{4\pi i}{3}}}{\sqrt{1+tan\ \theta}}|1\rangle\right) ,                   \\
&|\phi_3\rangle=\left(\frac{1}{\sqrt{1+tan\ \theta}}|0\rangle+\frac{1}{\sqrt{1+cot\ \theta}}|1\rangle\right) \otimes \left(\frac{1}{\sqrt{1+cot\ \theta}}|0\rangle+\frac{1}{\sqrt{1+tan\ \theta}}|1\rangle\right).
\end{split}
\end{equation}

\begin{table}[]
 \begin{tabular}{|c|c|c|c|c|c|c|c|}
\hline
\toprule
\multicolumn{2}{|c|}{Item}   & $P_{XX}^-$ &$P_{YY}^-$&$P_{ZZ}^-$ &  $|\phi_1\rangle\langle\phi_1|$ &$|\phi_2\rangle\langle\phi_2|$ & $|\phi_3\rangle\langle\phi_3|$ \\ \midrule
\hline
\multirow{2}{*}{Alice} & QWP & $45^\circ $& $0^\circ $& $0^\circ $& $30.1564^\circ $& $30.1564^\circ $& $ 52.9551 ^\circ $ \\ \cmidrule(l){2-8}
\cline{2-8}
                       & HWP &$22.5^\circ $& $67.5^\circ $& $0^\circ $& $45.9801^\circ $& $74.1763^\circ $& $ 26.4774^\circ $ \\ \midrule
                       \hline
\multirow{2}{*}{Bob}   & QWP &$4^\circ $& $0 ^\circ $& $0 ^\circ $& $59.8436 ^\circ $& $59.8436  ^\circ $& $37.0449 ^\circ $  \\ \cmidrule(l){2-8}
\cline{2-8}
                       & HWP &$22.5^\circ $& $67.5^\circ $& $0  ^\circ $& $60.8237 ^\circ $& $   89.0199^\circ $& $  18.5226 ^\circ $\\ \bottomrule
                       \hline
\end{tabular}
\caption{The angles of wave plates used to perform different measurements in the verification of pure two-qubit states.  }
\label{my-label}
\end{table}
 For each pair of the entangled states, one of the four measurements is randomly chosen with weight ($\frac{2-sin\ 2\theta}{4+sin\ 2\theta}$,$\frac{2(1+sin\ \theta)}{3(4+sin\ \theta)}$,$\frac{2(1+sin\ \theta)}{3(4+sin\ \theta)}$,$\frac{2(1+sin\ \theta)}{3(4+sin\ \theta)}$).  It is easy to check that the four probabilities add up to 1.
 With these measurement settings, $f(\theta)$ is calculated as $(2+\sin\theta\cos\theta)^{-1}$.

The concrete angles of HWP and QWP used to construct the corresponding measurement settings for verification are shown in Table I.

{}

{\bf Acknowledgments}

\textbf{Funding:} This work was supported by the National Key Research and Development Program of China (Nos. 2016YFA0302700, 2017YFA0304100), National Natural Science Foundation of China (Grant Nos. 11874344, 61835004, 61327901, 11774335, 91536219, 11821404), Key Research Program of Frontier Sciences, CAS (No. QYZDY-SSW-SLH003), Anhui Initiative in Quantum Information Technologies (AHY020100, AHY060300), the Fundamental Research Funds for the Central Universities (Grant No. WK2030020019, WK2470000026), Science Foundation of the CAS (No. ZDRW-XH-2019-1). W.-H.Z. and Z.C. contribute equally to this work.
\textbf{Author Contributions:} W.-H.Z. and G.C. planned and designed the experiment.
Z.C. and X.-J.Y. proposed the framework of the theory and made the calculations.
W.-H.Z. carried out the experiment
assisted by P.Y., J.-S.X., S.Y. and X.-Y.X., whereas X.-X.P. designed the computer
programs.
G.C. and Y.-J.H. analyzed the experimental
results and wrote the manuscript. G.-C.G. and C-F.L. supervised
the project. All authors discussed the experimental
procedures and results. \textbf{Competing interests:} The authors
declare that they have no competing interests. \textbf{Data and
materials availability:} All data needed to evaluate the
conclusions in the paper are present in the paper. Additional data related to
this paper may be requested from the authors.

\end{document}